\documentclass[sigconf]{acmart}
\renewcommand\footnotetextcopyrightpermission[1]{} 
\usepackage{enumitem}
\usepackage{booktabs}
\usepackage{multirow}
\usepackage{makecell}
\usepackage{threeparttable}
\acmConference[NTCIR-19]
{The 19th NTCIR Conference}
{Dec. 8--10, 2026}
{Tokyo, Japan}
\begin{document}

\title{Multi-Modal Agricultural Document Understanding at the DAGRI Task}


\author{Zhefan Wang}
\authornote{Equal contribution.}
\affiliation{%
  \institution{Shanghai Artificial Intelligence Laboratory}
  \institution{College of Computer Science and Artificial Intelligence, Fudan University}
  \country{China}
}
\email{wangzhefan@pjlab.org.cn}

\author{Da Li}
\authornotemark[1]
\affiliation{%
  \institution{State Key Laboratory of AI Safety}
  \institution{Institute of Computing Technology, Chinese Academy of Sciences}
  \institution{University of Chinese Academy of Sciences}
  \country{China}
}
\email{lida21s@ict.ac.cn}

\author{Keping Bi}
\authornote{Corresponding authors.}
\affiliation{%
  \institution{State Key Laboratory of AI Safety}
  \institution{Institute of Computing Technology, Chinese Academy of Sciences}
  \institution{University of Chinese Academy of Sciences}
  \country{China}
}
\email{bikeping@ict.ac.cn}

\author{Nanqing Dong}
\authornotemark[2]
\affiliation{%
  \institution{Shanghai Artificial Intelligence Laboratory}
  \institution{Shanghai Innovation Institute}
  \country{China}
}
\email{dongnanqing@pjlab.org.cn}

\begin{abstract}
Agricultural documents, such as prefectural technical reports and crop calendars, are critical knowledge resources. However, their complex table layouts, significant variations in document length, and dense multi-modal elements pose substantial challenges for automated understanding. To address these issues, the NTCIR-19 DAGRI (Data Analytics for aGRicultural Information) task introduced three subtasks focusing on different dimensions of document intelligence. In the Table Information Extraction (TableIE) task, we developed a multi-stage pipeline combining structure-preserving parsing via MinerU, region-adaptive segmentation for long documents, and a dual-modal (vision-text) extraction strategy for complex work schedules. This approach achieved first place with an exact match score of 0.6566. For Multi-Table Multi-Hop Question Answering (MTMH-QA) over interrelated JSON tables, we adopted an agentic search framework utilizing DCI-Agent to iteratively explore directories via bash commands, reinforced by a cross-model validation mechanism. This yielded a revised accuracy of 0.8444, initially ranked first. In the Figure and Table Question Answering (FiT-QA) task, which requires comprehending single large composite images, we implemented a dual-modal VLM pipeline enhanced by deterministic answer normalization, securing the highest BLEU score. Together, our system effectively addresses regional structural variations, dynamic cross-file evidence discovery, and multi-modal alignment, providing a robust framework for complex agricultural document understanding.

\end{abstract}


\keywords{Agricultural Document Understanding, Table Information Extraction, Multi-Hop Question Answering, Vision-Language Models}
\maketitle

\section*{Team Name}
Information Retrieval Group (irg88888)

\section*{Subtasks}
Table Information Extraction (TableIE), Multi-Table Multi-Hop Question Answering (MTMH-QA), Figure and Table Question Answering (FiT-QA)

\section{Introduction}

Agricultural documents, including crop cultivation guides, operational plans, and farm income and expenditure records, provide essential knowledge for agricultural production, farm management, policy analysis, and rural economic development~\cite{gauba2026agmmu}. These documents integrate diverse forms of information, including structured data presented in tables, visual content conveyed through charts, and narrative text comprising technical instructions. This diversity presents substantial challenges for automated information extraction and understanding, including extracting structured data from complex layouts, performing cross-table reasoning based on multi-source data, and comprehending composite content where text and visuals are intertwined.
To support systematic research into these challenges, NTCIR-19~\cite{dagri-1-overview} introduced the Data Analytics for Agricultural Information task, known as DAGRI, which comprises three subtasks, each focusing on a distinct aspect of agricultural document understanding. The first is Table Information Extraction (TableIE)~\cite{yang2025enhanced}, which requires the system to extract structured table data from agricultural PDF reports and convert it into a unified, machine-readable format. The second, Multi-Table Multi-Hop Question Answering (MTMH-QA)~\cite{zhu2021tat}, focuses on answering questions through multi-step reasoning over a corpus of JSON-formatted tables. The third is Figure and Table Question Answering (FiT-QA)~\cite{lompo2025visual}, which involves answering questions based on images containing tables and charts. 

For TableIE, the challenges arise from variations in table structure and document scale. 
Tables exhibit complex layouts characterized by diverse row and column arrangements, merged cells, and multi-level headers. Some complex tables contain a large number of rows and columns, while the relationships between data cells and their row or column headers depend heavily on spatial layout. In addition, agricultural documents vary significantly across regions in both length and table type, making unified processing difficult. 
To address these challenges, we design a multi-stage processing pipeline. First, we use MinerU~\citep{chen2024mineru}, an OCR parsing tool, to convert PDF documents into structured Markdown while preserving the original table structures. We then divide long documents into sub-documents by crop type to reduce the complexity of input. 
During the initial extraction stage, we combine keyword filtering with large language models (LLMs) to extract information from the OCR text of all tables. For layout-dependent complex tables such as work schedules, we further use vision-language models (VLMs) to jointly process the original image and its corresponding OCR text. VLMs use visual layout information to correct the mappings between data cells and their row or column headers, thereby improving extraction accuracy.


Whereas TableIE emphasizes document parsing, MTMH-QA requires systems to answer questions through multi-hop reasoning over a corpus of JSON-formatted tables. 
Questions based on these tables often involve precise numerical values, making exact localization essential. Their answers may also depend on evidence distributed across multiple tables, which must be retrieved and combined over successive reasoning steps. 
Given the small number of JSON files and the need for exact localization, we adopt DCI-Agent, a Direct Corpus Interaction framework for active exploration of the source files. At each step, the agent formulates and executes Bash commands based on the question and the evidence collected so far, examines the retrieved content, and uses new findings to determine the next search. This process enables direct evidence localization without relying on semantic search while preserving the reasoning chain. 
To improve stability, we run the agent independently using two different models and synthesize the results through an arbitration strategy to mitigate the inherent uncertainty of a single run.

FiT-QA is a question-answering task based on images. It requires systems to answer questions using a single image that contains tables, charts, and dense textual annotations. The difficulties arise from visual content understanding and precise answer generation. First, systems must integrate visual information with textual semantics to establish relationships among different information elements. Second, the task imposes strict constraints on answer format. Even when an answer is semantically correct, differences in numerical formatting, unit notation, or punctuation may affect the final score. 
We therefore develop a two-stage pipeline using MinerU and VLMs. First, we use MinerU to extract text from the image and provide both the original image and the extracted text to the vision-language model for question answering. And then, we apply a normalization procedure to produce the answer in the required format.

Overall, our optimizations yielded competitive results in the DAGRI Formal Run.
In Table IE, we ranked first with a final score of 0.6566, demonstrating the cross-regional adaptability of region-adaptive document processing and structure-preserving parsing. 
Performance on MTMH-QA was similarly strong: the system initially secured first place with an accuracy of 0.7350, which increased to 0.8444 after questions with erroneous or ambiguous reference answers were excluded. 
For FiT-QA, our system obtained the highest BLEU score and ranked second according to both the LLM Judge and human evaluation. 
The gap between BLEU and the LLM-based and human evaluations suggests that answer normalization improves alignment with reference answers but cannot compensate for errors in visual interpretation or missing information in the generated answers.
Overall, these results demonstrate the effectiveness and generalizability of our framework for agricultural document understanding.



\section{Related Work}
\label{sec:related}
This section reviews related work in three areas: table information extraction, agentic search for multi-hop question answering, and visual document question answering.
\subsection{Table Information Extraction}
Extracting structured information from tables in PDFs is a fundamental task in document understanding, with key challenges arising from complex table layouts, cross-modal alignment, and fine-grained boundary detection~\cite{zanibbi2004survey,embley2006table}. Rule-based systems typically infer table structures from text coordinates, whitespace, and ruling lines, but their dependence on handcrafted assumptions limits their robustness across heterogeneous layouts. Neural models~\cite{schreiber2017deepdesrt,paliwal2020tablenetdeeplearningmodel} formulate table extraction as a visual detection or segmentation problem. 
Layout-aware pre-trained models further integrate textual content with two-dimensional spatial information~\cite{lo2019layoutlm,xu2021layoutlmv2,huang2022layoutlmv3}. In contrast, specialized models and datasets, including PubTabNet~\cite{zhong2020pubtabnet} and Table Transformer~\cite{smock2022pubtables}, support the recovery of cell boundaries and logical structures.
Nevertheless, recognizing individual cells does not necessarily recover tables’ semantic organization, particularly when headers are hierarchical, units appear outside the table body, or values are associated across distant regions. These challenges are especially pronounced in agricultural reports, which frequently contain merged cells, sparse calendar-style grids, dense annotations, and tables spanning multiple pages. Linear extraction tools such as pdftotext flatten these layouts into text sequences, disrupting critical associations among headers, values, units, and annotations. Reliable extraction therefore requires preserving both the geometric structure and the semantics of the original tables. 


\subsection{Agentic Search}
Agentic search enables LLMs to alternate between reasoning and information retrieval, adjusting subsequent actions based on previously obtained results. ReAct~\cite{yao2023react} interleaves reasoning traces with actions, allowing LLMs to retrieve external information while updating their action plans. 
Search-o1~\cite{li2025searcho1} dynamically invokes retrieval when knowledge gaps arise and refines retrieved documents before incorporating them into the reasoning process. Search-R1~\cite{jin2025searchr1} uses reinforcement learning to train LLMs to generate multiple queries during step-by-step reasoning autonomously. For multi-hop question answering, ChainRAG~\cite{zhu2025chainrag} progressively rewrites queries using entities identified in previous hops, while S2G-RAG~\cite{li2026s2grag} explicitly evaluates evidence sufficiency and converts unresolved information gaps into subsequent queries. Agentic-R~\cite{liu2026agenticr} further adapts the retriever to multi-turn search by jointly considering passage relevance and final-answer correctness. These methods improve over static RAG pipelines by dynamically refining queries, evidence, and stopping decisions. This paradigm is particularly suitable for MTMH-QA, where evidence must be iteratively discovered and analyzed across multiple tables.


\begin{figure*}[htbp]
    \centering
    \includegraphics[width=0.8\linewidth]{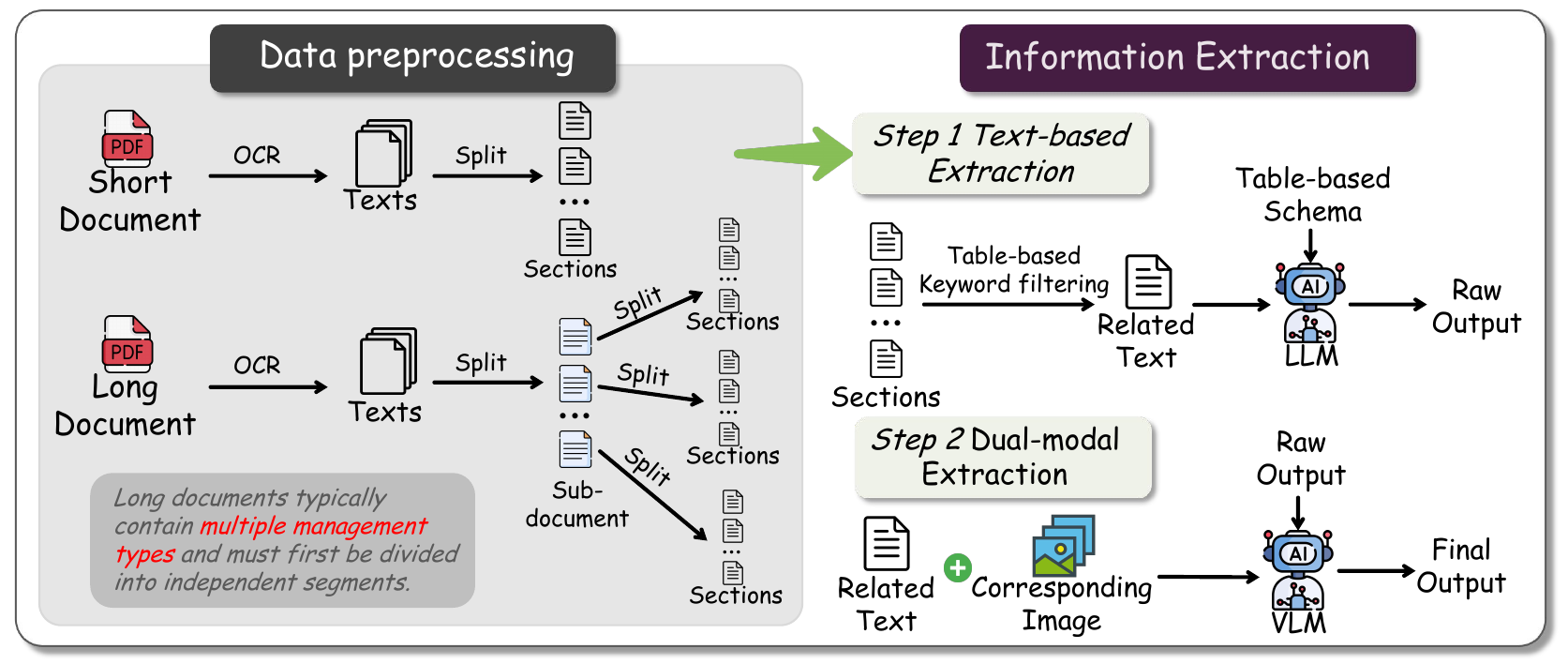}
    \caption{Multi-stage processing pipeline for the TableIE task. The preprocessing segments long documents into independent sub-documents based on management types. The extraction uses text-based extraction for general tables, while complex tables such as work schedules are processed by VLMs that jointly analyze images and OCR text.}
    \label{fig:TABLEIE_main_fig}
\end{figure*}

\subsection{Visual Document Question Answering}
Vision-language models (VLMs) have achieved remarkable performance on visual document understanding tasks. 
Several benchmarks have been proposed to evaluate VLMs on chart and figure understanding. ChartQA~\cite{kafle2022chartqa} introduced a benchmark for chart question answering that includes both visual and logical reasoning challenges, revealing that existing models struggle with tasks requiring visual perception of chart elements such as reading values from bar charts as well as multi-step reasoning over chart data. DocVQA~\cite{mishra2022docvqa} extended the VQA paradigm to images of documents, requiring VLMs to answer questions based on  complex visual layouts. More recently, mPLUG-DocOwl~\cite{hu2024mplug} proposed a modularized multimodal large language model for visual document understanding that combines a vision encoder with a language model, achieving strong results on document-level QA tasks. 
While models like Donut~\cite{lee2023donut} demonstrated that end-to-end document understanding without explicit optical character recognition (OCR) is possible, they often misread small text, dense cells, or domain-specific chart annotations. From LLaVA~\citep{llava} to the latest Qwen~\citep{bai2025qwen3}, VLMs have significantly improved their ability to process high-resolution images. This progress enables more accurate recognition of small text, chart elements, and complex document layouts, enhancing VLMs’ ability to understand visual documents.
Because tables are presented as images, TableIE and FiT-QA can both be viewed as forms of visual document question answering.


\section{Methods}
\label{sec:methods}
This section presents our optimizations for three subtasks.
\subsection{TableIE}

\begin{figure}[htbp]
  \centering
  \includegraphics[width=0.65\columnwidth]{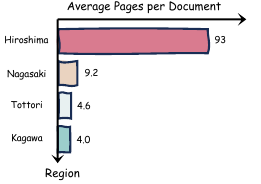}
  \caption{Comparison of average document length across regions. The test set comprises a total of 27 documents. The extreme variance in page counts necessitates our region-adaptive segmentation strategy.}
  \label{fig:page_variance}
\end{figure}

The TableIE task requires extracting Management Types (MT) and Management Indicators (MI) from Japanese agricultural PDFs. Based on our analysis of the dataset, the challenges can be summarized across two dimensions. First, table layouts possess inherent complexity, featuring diverse row/column structures, merged cells, and multi-level headers, all of which must be faithfully preserved during parsing as they carry critical semantic information. Second, the length and structure of agricultural documents vary significantly across regions (summarized in Figure \ref{fig:page_variance}), as does the complexity of different table categories. Documents from Hiroshima can span dozens to hundreds of pages and contain multiple management types, while documents from Nagasaki and Tottori are typically shorter, with each containing only a single management type. Furthermore, different table categories exhibit varying complexity: complex tables such as work schedules often span over twenty rows and thirty columns, where the correspondence between data cells and row/column headers is highly dependent on visual positioning, whereas simple tables such as income/expenditure summaries are much smaller and more straightforward (see Figure 3). We design a multi-stage processing pipeline, as illustrated in Figure ~\ref{fig:TABLEIE_main_fig}.

\begin{figure}[htbp]
    \centering
    \includegraphics[width=0.95\linewidth]{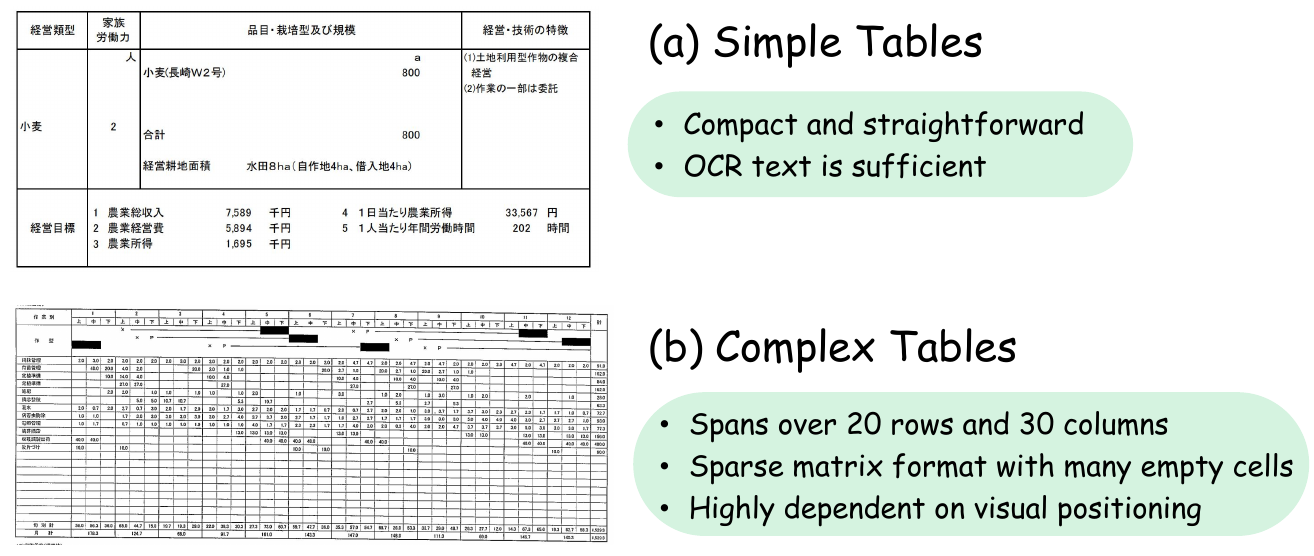}
    \caption{Comparison of table complexity. (a) Simple tables are compact and straightforward, where OCR text is sufficient. (b) Complex work schedules feature a sparse matrix format and heavily depend on visual positioning, necessitating dual-modal extraction.}
    \label{fig:table_example}
\end{figure}

\textbf{Structure-preserving Document Parsing.} 
To address the challenge of complex table layouts, we employ MinerU, a high-throughput document parsing tool, for structure-preserving OCR during the data preprocessing stage. MinerU uses a locally deployed VLM backend to parse agricultural PDFs and recover tables as structured HTML. The extracted HTML preserves merged cells through \texttt{rowspan} and \texttt{colspan} attributes, retains multi-level header hierarchies, and maintains alignments between headers and data cells. This structured representation can be directly processed as text, reducing multimodal input overhead. 

\textbf{Region-adaptive Segmentation and Text-based Extraction.} 
To address variation across regions, we adopt region-adaptive processing strategies. To reduce complexity, we use regular expressions to hierarchically segment long documents into management-type sub-documents, which are further divided into sections by numbered headings. 
For general tables, the system performs text-based extraction (Step 1 in Figure \ref{fig:TABLEIE_main_fig}): for each sub-document, the system extracts relevant sections containing predefined keywords for target tables, and feeds this filtered content to the LLM along with a type-specific schema, which incorporates field-level constraints like unit conversion rules, period format normalization, and nested definitions to ensure consistent structured outputs.
This approach reduces the input length per invocation and isolates extraction failures. For shorter documents containing only a single management type, no segmentation is needed. Before extraction, however, two LLM-based classifiers are introduced: the first returns an empty result when no relevant tables exist to prevent hallucinated extraction, while the second determines whether income/expenditure data belong to the MT or MI level, ensuring extraction occurs only at the appropriate level. To identify MIs across multiple segments, an LLM scans the entire document, combining information like variety prefixes, names, and cultivation types to construct complete crop names, treating each as an independent MI.

\textbf{Dual-modal Extraction for Complex Tables.} 
To address the challenge of complex table types, we adopt a dual-modal extraction strategy specifically for complex tables (Step 2 in Figure \ref{fig:TABLEIE_main_fig}). During the extraction process, we observed that work schedule tables have significantly lower extraction accuracy than other table types, as their sparse matrix format with many empty cells complicates text-only row/column alignment. When using text-only extraction, the VLM struggles to correctly determine which data cell belongs to which row or column. To address this, for work schedule tables, we simultaneously feed the corresponding page images and Markdown text into the VLM, with the prompt explicitly instructing the model to leverage both image and text modalities and to prioritize the visual table structure. This dual-modal input enables the VLM to utilize visual positional relationships within the tables to correctly map data to the corresponding row and column headers. For other table types such as income/expenditure tables and cultivation area tables, which are smaller and less complex, we use single-modal text extraction only, as the OCR text is sufficient for accurate extraction.

\subsection{MTMH-QA}
Multi-Table Multi-Hop Question Answering requires systems to answer questions about agricultural operations by locating and integrating relevant information from a corpus containing hundreds of tables. The corpus covers municipalities, crops, management types, income and expenditure, cultivation scale, work schedules, equipment, and labor. 
Questions posed over this corpus typically require multi-hop reasoning across files, including locating exact values in specific tables, associating fields across files, filtering records, aggregating results, and performing numerical calculations. A substantial proportion of the questions also involve localized exact matching, requiring systems to identify specific fields or values within particular tables. 
Relevant evidence is often distributed across files, and the next file to inspect may become apparent only after an intermediate result is obtained. Many questions also depend on exact field names, entities, or numerical values, making lexical matching more important than semantic matching. 
To support adaptive exploration and precise evidence localization, we adopt agentic search, which enables LLMs to inspect the corpus iteratively, refine subsequent searches based on intermediate findings, and validate the evidence collected at each step. This process allows them to locate exact values across files and maintain the evidence chain required for multi-hop reasoning.

\begin{figure*}[htbp!]
    \centering
    \includegraphics[width=0.85\linewidth]{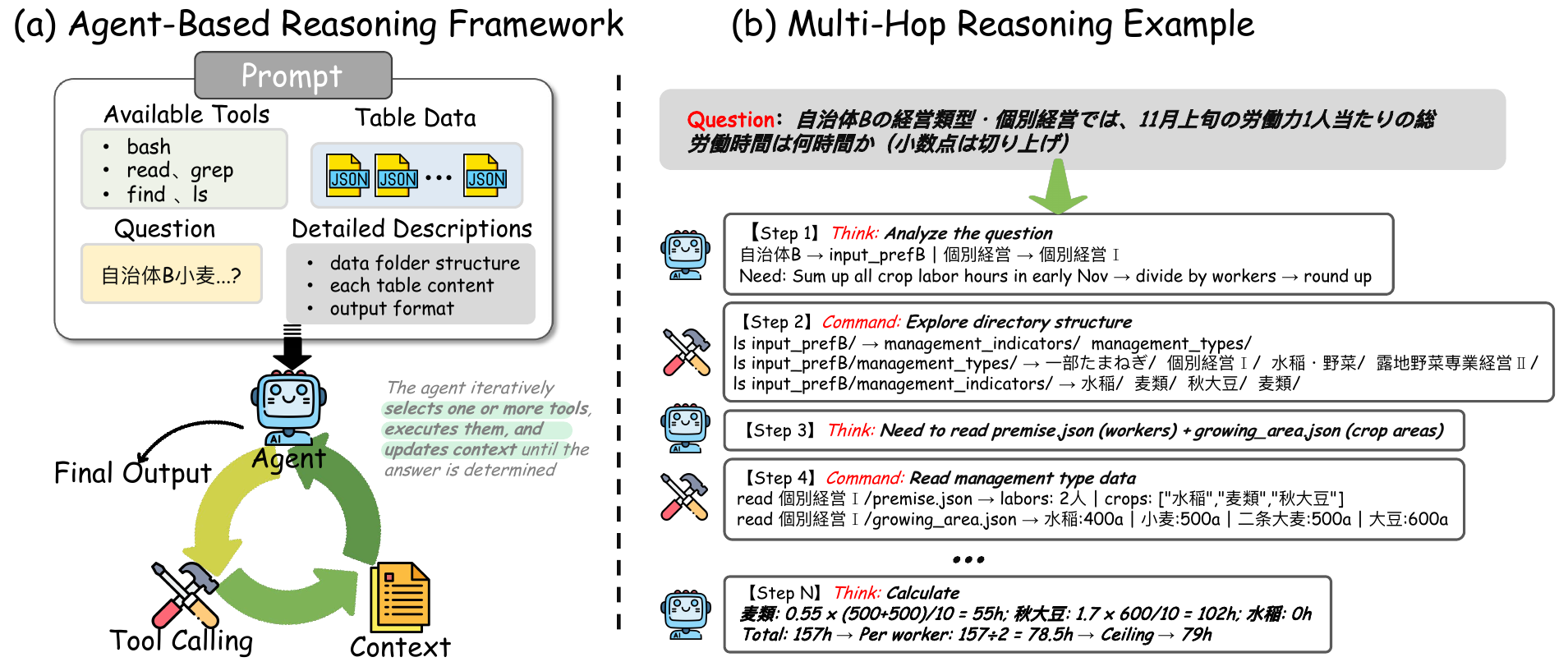}
    \caption{Illustration of DCI-Agent. (a) Agent iteratively interacts with tabular data using shell tools. (b) A concrete example of how the agent performs multi-hop searches to obtain the final answer.}
    \label{fig:MTMH_main_fig}
\end{figure*}

\textbf{DCI Agent.} Based on these characteristics, we enable LLMs to search the corpus through direct interaction rather than semantic matching. Agents following this approach are referred to as DCI-Agents. LLMs are allowed to invoke shell tools such as \texttt{grep}, \texttt{find}, and \texttt{ls} to locate exact numerical values within JSON files, shown in Figure ~\ref{fig:MTMH_main_fig}(a). 
For each question, we provide agents with the data directory structure, JSON schemas, and field relationships. This information defines the evidence search space and explains the semantic connections among files. Agents use \texttt{find} and \texttt{ls} to identify relevant directories and candidate files, \texttt{grep} to locate specific entries, and \texttt{read} to retrieve specific table contents. For questions requiring filtering, aggregation, or numerical calculation, the agent can compose these commands to process and verify the collected data.
Each tool result guides the next search or reasoning step. When the available evidence is insufficient, the agent expands its search; when information from multiple files must be connected, it uses schema-defined field relationships to acquire additional evidence. 
The Agent terminates the search and produces a final answer only when the evidence sufficiently supports the required answer. 
Through this iterative process, agents can construct and validate cross-file evidence chains and discover information that becomes relevant only after intermediate reasoning steps. Figure ~\ref{fig:MTMH_main_fig}(b) illustrates a concrete multi-hop reasoning example of this step-by-step exploration.

\textbf{Cross-Agent Validation.} 
Agent outputs may contain errors in evidence interpretation, reasoning, or calculation. When the system relies on a single agent, it lacks an independent signal to identify these errors, and repeated inference may reproduce them rather than reveal them. To reduce this risk, we use GPT-5.5~\footnote{\url{https://openai.com/index/introducing-gpt-5-5}} and Qwen3.7-Max~\footnote{\url{https://qwenlm.github.io}} as independent agents. For each question, we compare the independently generated answers using exact matching. Questions with different responses are treated as requiring additional verification and are re-evaluated using Qwen3.7-Max, which is chosen because it generally produces concise answers that better conform to the expected numerical format, whereas GPT-5.5 tends to generate more detailed or conditional responses. By applying additional inference only to model-sensitive cases, we reduce the influence of single-model errors while limiting unnecessary computational overhead.

\subsection{FiT-QA}





The FiT-QA task requires answering questions about agricultural documents with complex layouts, dense information, and domain-specific terminology. 
The input for each question is a single large composite image containing multiple tables, embedded line or Gantt charts, and fine-grained textual and numerical annotations, requiring joint understanding of visual layouts and text semantics.
The task presents two main challenges. 
First, VLMs must accurately locate regions within images that feature vivid colors and dense information, recognize small text and values, and summarize and reason over multiple visual elements.
Second, model outputs may differ stylistically from reference answers through vague wording, unnecessary annotations, or missing units.
Such stylistic mismatches lead to substantial score drops on surface-form metrics such as bilingual evaluation understudy (BLEU), even when semantically correct.

\begin{figure}[htbp!]
    \centering
    \includegraphics[width=0.75\linewidth]{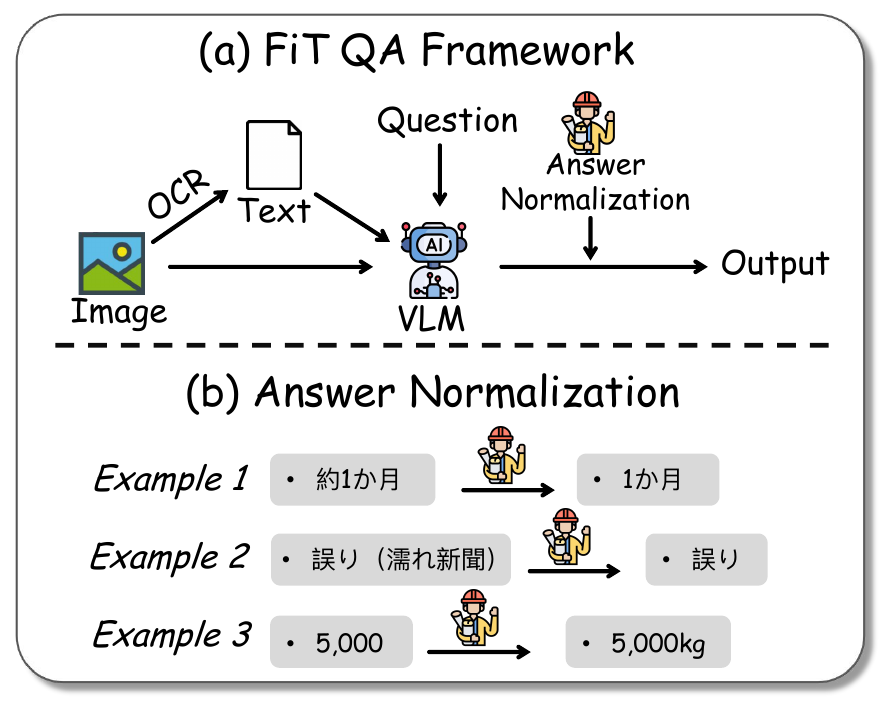}
    \caption{Our optimizations for FiT-QA. (a) A question-answering pipeline integrating OCR text and original images. (b) Examples of the deterministic answer normalization process, which eliminates stylistic discrepancies to align with reference answers.}
    \label{fig:FIT_main_fig}
\end{figure}

\textbf{Dual-modal understanding.} To address the problem of insufficient visual perception, we design a two-stage dual-modal pipeline, as shown in Figure ~\ref{fig:FIT_main_fig}(a). 
We first process each PNG image with MinerU to recover table structures and textual content as structured Markdown. For question answering, we provide VLMs with both the original image and the OCR-generated Markdown, allowing them to combine visual evidence with structured textual information.
This design ensures the two modalities complement each other: when OCR misses part of the content, VLMs can refer directly to visual cues in the original image for supplementary understanding; when precise reading of text from small or dense cells is required, the OCR text serves as an anchor providing structured textual information, preventing VLM visual hallucination. This dual-modal input enables the system to simultaneously leverage the structured textual information provided by OCR and the visual context from the original image, significantly improving comprehensive understanding of agricultural images with dense annotations, fine-grained scales, and multi-panel layouts.

\textbf{Post-processing pipeline.} 
To address these stylistic mismatches, we therefore design a deterministic post-processing pipeline, as illustrated in Figure~\ref{fig:FIT_main_fig}(b), that eliminates stylistic differences without altering the underlying information content. 
The pipeline comprises three steps. First, vague expression removal: redundant qualifiers such as ``approximately or so'' and similar hedging phrases are stripped, retaining only the numerical value itself. Second, polite-to-plain form conversion: the polite forms that VLMs tend to output are uniformly converted to the plain forms commonly used in written Japanese, eliminating stylistic inconsistency with the reference answers. Third, unit completion: when the question explicitly requires a specific unit (e.g., ``how many cm'') but the answer provides only a numerical value, the corresponding unit is automatically appended based on the question context, converting ``30'' to ``30cm''. Additionally, the pipeline handles edge cases such as merging parenthetical annotations, effectively eliminating systematic score loss caused by stylistic differences.



\section{Experiments}
\label{sec:experiments}
We evaluate our optimization on the three subtasks under the official formal run setting. This section first describes configurations used in our experiments, followed by the results for each subtask.

\subsection{Experimental Setup}
All experiments are conducted on the official DAGRI formal run datasets~\cite{dagri_overview}. The model configurations for each subtask are as follows:
\begin{itemize}[leftmargin=*]
    \item \textbf{TableIE}: 
    We use Qwen3.7-Max for text-based structured information extraction and final answer generation, while utilizing GPT-5.5 as the VLM for the dual-modal extraction of complex tables.
    \item \textbf{MTMH-QA}: We employ GPT-5.5 and Qwen3.7-Max as two independent agents. When their answers differ, Qwen3.7-Max performs an arbitration rerun. Both models are accessed through compatible API interfaces.
    \item \textbf{FiT-QA}: We use GPT-5.5 for answer generation and set the temperature to 0.0 to ensure deterministic outputs.
\end{itemize}
We deploy MinerU locally through its API using the VLM auto engine. OCR processing runs with a concurrency level of 32, and MinerU outputs are pre-computed and cached for all subtasks. Each MTMH-QA agent is allowed up to 100 turns and 6,000 seconds per question, with model-specific concurrency limits to prevent rate limiting. If an agent returns an empty answer, the process is retried up to three times using a clean output directory for each attempt.

\subsection{TableIE Results}
Table~\ref{tab:tableie_results} summarizes the exact match accuracy for TableIE at the field level. We evaluate the proposed method on the training set, which contains documents from Hiroshima, Nagasaki, Tottori, and Kagawa, and report the performance of the final submission on the test set.


\begin{table}[htbp]
\centering
\caption{Field-level exact match accuracy on the TableIE task. The final results incorporate dual-modal extraction optimization for work schedule tables.}
\label{tab:tableie_results}
\resizebox{0.7\linewidth}{!}{
\begin{tabular}{lcc}
\toprule
\textbf{Evaluation Scope} & \textbf{Train Set} & \textbf{Test Set} \\
\midrule
\multicolumn{3}{l}{\textit{\textbf{Official Baseline}}} \\
\textit{Nagasaki and Tottori} & 0.0953 & - \\
\textit{Hiroshima} & 0.0 & - \\
\textit{Overall} & 0.0576 & - \\ \midrule
\addlinespace
\multicolumn{3}{l}{\textit{\textbf{Ours (w/o Dual-modal)}}} \\
\textit{Nagasaki and Tottori} & 0.6005 & - \\ 
\textit{Hiroshima} & 0.6474 & - \\
\textit{Overall} & 0.6207 & - \\ \midrule
\addlinespace 
\multicolumn{3}{l}{\textit{\textbf{Ours}}} \\
\textit{Overall} & - & \textbf{0.6566} \\
\bottomrule
\end{tabular}}
\end{table}
The results reveal noticeable performance variation across regions, with better extraction performance on Hiroshima documents than on Nagasaki and Tottori documents. Our proposed method also maintains consistent performance across the training and testing sets, demonstrating its ability to handle diverse regional document formats.


\subsection{MTMH-QA Results}

\begin{table}[htbp!]
  \centering
  \caption{MTMH-QA Performance Results.}
  \label{tab:mtmh_performance}
  \begin{tabular}{lc}
    \toprule
    Method & Score \\
    \midrule
    Baseline & 0.1682 \\
    DCI + Qwen3.7-Max & 0.7115 \\
    DCI + GPT-5.5 & 0.7158 \\
    DCI + Cross-Agent Validation & \textbf{0.7350} \\
    \bottomrule
  \end{tabular}
\end{table}
Table \ref{tab:mtmh_performance} summarizes the overall performance of the models in the MTMH-QA pipeline.
GPT-5.5 and Qwen3.7-Max answer each question independently. Identical answers are accepted directly, while inconsistent answers are forwarded to the arbitration stage.


\begin{table}[htbp!]
  \centering
  \caption{Model Agreement Statistics in MTMH-QA.}
  \label{tab:mtmh_agreement}
  \begin{tabular}{lc}
    \toprule
    Category & Count \\
    \midrule
    Total Questions & 468 \\
    Exact Agreement & 378 (80.8\%) \\
    Disagreement & 90 (19.2\%) \\
    \bottomrule
  \end{tabular}
\end{table}

As detailed in Table \ref{tab:mtmh_agreement}, the two agents produced inconsistent answers for 90 out of 468 questions (19.2\%).
We manually analyze the 90 disagreements and identify four recurring sources:
\begin{itemize}[leftmargin=*]
    \item \textbf{Numerical Precision.} 
    Many disagreements arise from different rounding strategies or decimal precision. For example, GPT-5.5 produces ``166666 yen'' and ``12.1h'', whereas Qwen3.7-Max produces ``166000 yen'' and ``12.11h''. These differences indicate inconsistent handling of intermediate calculations and numerical formatting.
    \item \textbf{Answer Format.} 
    Some disagreements arise from differences in output style rather than the underlying conclusion. For example, one agent may answer ``No'', while the other uses ``Impossible''. Similar variations occur when the same result is expressed as structured JSON by one agent and as plain text by the other. These cases indicate inconsistent adherence to the required output format.
    \item \textbf{Multi-file Reasoning.} 
    Substantial disagreements occur when questions require evidence from several files, such as calculating marginal profit from balance and cultivation area data. The agents may select different evidence or follow different intermediate reasoning steps, leading to inconsistent results.
    \item \textbf{Unit Interpretation.} 
    There are some disagreements resulting from unit interpretation or conversion, such as ``4L'' versus ``40L''. Others are purely formatting differences, such as ``1,200kg'' versus ``1200kg''. The latter represents equivalent answers that are nevertheless treated as different under exact match evaluation.
\end{itemize}

The complementary response patterns of GPT-5.5 and Qwen3.7-Max contribute to the effectiveness of cross-model validation. The arbitration results further reduce inconsistencies and produce more accurate and standardized final answers.

\subsection{FiT-QA Results}
We evaluate our pipeline using BLEU and LLM judge scores, as reported in Table~\ref{tab:fitqa_results}. We conduct an ablation study on the training set to examine the contributions of OCR and post-processing, and evaluate the complete configuration on the testing set.

\begin{table}[htbp!]
\centering
\caption{Ablation results on FiT-QA. All configurations use the same model.  The first row is the baseline.}
\label{tab:fitqa_results}
\small
\resizebox{0.9\linewidth}{!}{
\begin{tabular}{ccccc}
\toprule
\textbf{Split} & \textbf{OCR} & \textbf{Post-processing}
& \textbf{BLEU} & \textbf{LLM Judge} \\
\midrule
\multirow{3}{*}{\textit{\textbf{Train}}}
& $\times$  & $\times$  & 39.2816 & 4.1796 \\
& $\checkmark$ & $\times$  & 51.7340 & \textbf{4.3169} \\
& $\checkmark$ & $\checkmark$ & \textbf{58.1943} & 4.3055 \\
\midrule
\textit{\textbf{Test}}
& $\checkmark$ & $\checkmark$ & 54.5788 & 3.9959 \\
\bottomrule
\end{tabular}
}
\end{table}

The results on the training set demonstrate the contribution of each component. Incorporating MinerU OCR substantially improves both BLEU and LLM judge scores because the structured OCR text provides reliable textual evidence for small labels and dense table cells, while the original image preserves the visual layouts and relationships among chart elements.
Building on these gains, Post-processing further improves the BLEU score, while the LLM judge score remains relatively stable. This suggests that post-processing primarily improves alignment with reference answers by standardizing expressions, completing missing units, and removing redundant annotations without substantially changing the semantic content.
We further evaluate the complete configuration on the testing set. Although its performance is slightly lower than that on the training set, possibly due to differences in document complexity and question distribution, the results indicate that the proposed pipeline remains effective on unseen samples. Taken together, combining visual information with OCR text improves content understanding, while postprocessing increases the consistency of generated answers and their alignment with the expected answer format.


\section{Discussion}
\label{sec:discussion}

Our results across the three subtasks lead to three main observations. First, current LLMs still have limited ability to directly process long, dense, and structurally complex inputs. Auxiliary processing is therefore essential for improving content understanding. Structure-preserving OCR retains table hierarchies and cell-to-header associations, enabling LLMs to interpret values within the correct semantic context. Hierarchical segmentation reduces input length and irrelevant information, allowing the model to focus on coherent document sections and avoid context loss. Standardized answer parsing then converts free-form generations into the required schema, reducing ambiguity and improving output consistency.

Second, when evidence is not explicitly annotated, agentic search enables LLMs to locate relevant files, connect evidence across documents, and derive answers through iterative reasoning. This provides a practical alternative to fixed context selection for MTMH-QA. However, the efficiency and effectiveness of this approach depend partly on how the corpus is organized. Clear directory structures and consistent data schemas can guide the search process and reduce unnecessary exploration, whereas poorly organized corpora may increase retrieval costs and make relevant evidence more difficult to locate.


Finally, the divergence between BLEU and LLM judge scores highlights a broader issue in evaluation: surface-form metrics and semantic metrics capture different aspects of answer quality. Post-processing improved BLEU substantially while leaving the LLM judge score relatively stable, suggesting that style normalization primarily improves format alignment rather than reasoning quality. This finding underscores the need for evaluation frameworks that jointly consider both dimensions.

Insights from MTMH-QA Evaluation Dynamics: The ranking shift observed in the revised MTMH-QA evaluation provides critical insights into the limitations of purely retrieval-based reasoning. While our agentic search robustly handled standard multi-hop numerical queries (ranking first initially), the revised evaluation revealed specific vulnerabilities. According to the task organizers, systems frequently failed on ``trick questions'' that required recognizing that a requested cost item was not applicable or missing from the source tables altogether. Resolving these edge cases correctly requires determining that no additional calculations should be performed. The fact that Team NARO~\cite{dagri_overview}, utilizing Gemma4-31B, slightly outperformed our system on the cleaned dataset suggests that navigating these implicit conditions heavily relies on the base model's internal domain knowledge of Japanese farm management and agricultural accounting. This indicates that future agentic frameworks for agriculture must integrate external domain knowledge bases to better handle absent or inapplicable data fields.

\section{Conclusion and Future Work}
We presented our optimizations for the three NTCIR-19 DAGRI subtasks, combining structure-preserving parsing for TableIE, agent-based evidence exploration for MTMH-QA, and OCR-enhanced visual understanding for FiT-QA. 
The competitive results across all three subtasks demonstrate the effectiveness of document segmentation, agentic search, multimodal input, and answer normalization. Our findings also highlight the importance of preserving structural and visual information when processing complex agricultural documents.
Future work will address the broader challenge of enabling language models to reason reliably over heterogeneous, semi-structured documents. This requires bridging the gap between document perception and knowledge reasoning through unified representations of textual, visual, and structural information. We further aim to develop evidence-grounded reasoning mechanisms that can dynamically acquire information, trace intermediate decisions, and identify uncertainty or unsupported conclusions. These advances will support more reliable and transferable document intelligence systems.

\section{Limitations}
Despite the competitive results across the three subtasks, several limitations should be acknowledged. First, the experiments are conducted on a single benchmark with limited regional and document variation, leaving cross-domain and cross-lingual generalizability uncertain. Second, the contribution of individual components cannot be fully isolated because comprehensive ablation studies are unavailable. The reliance on large proprietary models also raises concerns about reproducibility, computational efficiency, and sensitivity to model updates. Finally, current evaluation metrics do not fully measure evidence faithfulness or reasoning validity, meaning that strong benchmark performance may not always correspond to reliable and auditable answers.


\begin{acks}
This work was supported by the National Natural Science Foundation of China (NSFC) under Grant No. 62302486, and the New Generation Artificial Intelligence National Science and Technology Major Project of China under Grant No. 2025ZD0121801.
\end{acks}

\bibliographystyle{ACM-Reference-Format}
\bibliography{ntcirsample}

\end{document}